\pdfoutput=1
\documentclass[journal=jpcafh, manuscript=article, layout=traditional]{achemso}

\usepackage{amsmath}
\usepackage{physics}
\usepackage{bbold}
\usepackage{amssymb}
\usepackage{bm}
\usepackage{verbatim}
\usepackage{graphicx}
\usepackage{url}
\usepackage{cancel}
\usepackage{siunitx}
\usepackage{appendix}
\usepackage[dvipsnames]{xcolor}

\renewcommand{\i}{{i}}
\newcommand{\f}{{f}}
\newcommand{\I}{{I}}
\newcommand{\F}{{F}}
\newcommand{\FC}{\text{FC}}
\newcommand{\FCHT}{\text{FCHT}}
\newcommand{\HT}{\text{HT}}
\newcommand{\T}{{T}}
\newcommand{\noZPE}{\text{noZPE}}

\SectionNumbersOn

\title{On the Importance of Well-Defined Thermal Correlation Functions in Simulating Vibronic Spectra}

\author{Rami Gherib}
\affiliation{OTI Lumionics Inc., 3415 American Drive Unit 1, Mississauga, Ontario L4V 1T4, Canada}
\email{rami.gherib@otilumionics.com}

\author{Scott N. Genin}
\affiliation{OTI Lumionics Inc., 3415 American Drive Unit 1, Mississauga, Ontario L4V 1T4, Canada}

\author{Ilya G. Ryabinkin}
\affiliation{OTI Lumionics Inc., 3415 American Drive Unit 1, Mississauga, Ontario L4V 1T4, Canada}
\email{ilya.ryabinkin@otilumonics.com}

\begin{document}

\maketitle

\begin{abstract}
  Two difficulties associated with the computations of thermal
  vibrational correlation functions are discussed. The first one is
  the lack of a well-behaved expression that is valid at both
  high-temperature and $T \rightarrow \SI{0}{\kelvin}$ limits.
  Specifically, if the partition function and the propagator are
  considered separately, then thermal vibrational correlation
  functions may have an indeterminate form $0/0$ in the limit
  $T \rightarrow \SI{0}{\kelvin}$. This difficulty is resolved when
  the partition function and the propagator are jointly considered in
  the harmonic approximation, which allows a problematic term that
  emanates from the zero-point energy to be cancelled out thereby
  producing a thermal correlation function with a determinate form in
  $T \rightarrow \SI{0}{\kelvin}$ limit. The second difficulty is
  related to the multivaluedness of the vibrational correlation
  function. We show numerically that an improper selection of branch
  leads to discontinuities in the computed correlation function and an
  incorrect vibronic spectra. We propose a phase tracking procedure
  that ensures continuity of both real and imaginary parts of the
  correlation function to recover the correct spectra. We support our
  findings by simulating the UV-vis absorption spectra of pentacene at
  \SI{4}{\kelvin} and benzene at \SI{298}{\kelvin}. Both are found to
  be in good agreement with their experimental counterparts.
\end{abstract}

\section{Introduction}
\label{sec:introduction}

\textit{In silico} design of light emitting materials is one of the
major applications of computational spectroscopy. Reliable predictions
of line positions and shapes in fluorescence and phosphorescence
spectra are in demand for identifying promising molecules among
thousands and sometimes tens of thousands candidates.

Simulations of vibrationally resolved electronic spectra can be
broadly classified as time-independent (TI)~\cite{ruhoff1994recursion,
  dierksen2005efficient, santoro2007effective, chang2008new} or
time-dependent (TD)~\cite{heller1981semiclassical, niu2010theory,
  peng2010vibration, baiardi2013general, de2018theoretical,
  beguvsic2020fly}. In TI approaches, spectra are simulated as
superposition of individual vibronic transitions, often broadened
phenomenologically to match experiments. TI methods become impractical
for large molecules due to the prohibitively large number of
contributing vibronic transitions~\cite{jacquemin2011td}. The
extensive computational costs have even prompted some researchers to
tackle the problem via quantum computing, particularly via boson
sampling and to rethink the computational hardware that ought to be
used for such problems~\cite{huh2015boson}. In TD approaches, vibronic
spectra are computed by the Fourier transform of the transition dipole
correlation function.

When vibronic transitions are considered, TI and TD formalisms require
at least two potential energy surfaces (PESs) --- one for the initial
and another for the final electronic state. Because the construction
of fully anharmonic multidimensional PESs is challenging for all but
the smallest molecules~\cite{ryabinkin2018qubit}, numerous TI and TD
approaches rely on the harmonic approximation (HA) in which the true
PESs are approximated by multidimensional paraboloids that are
constructed from truncated Taylor expansions at relevant nuclear
configurations. The relative position and orientation of two
paraboloids can be defined through the first-order Duschinsky
transformation~\cite{duschinsky1937importance}. Under the umbrella of
the HA, additional approximations can be made depending on the choice
of reference nuclear geometries where Taylor expansion is made. For
example, an adiabatic Hessian model evaluates the final electronic
state Hessian at a minimum of the final electronic state, whereas in a
vertical Hessian model, the minimum of the initial electronic state is
used~\cite{domcke1977comparison, hazra2004first, ferrer2012comparison,
  cerezo2013harmonic}. Further approximations, such as the independent
mode displaced harmonic approximation~\cite{petrenko2012efficient},
neglect Duschinsky rotation and/or assume that the initial and final
states have identical harmonic frequencies.

Numerous studies~\cite{niu2010theory, peng2010vibration,
  baiardi2013general, de2018theoretical, cerezo2013harmonic,
  ianconescu2004photoinduced, baiardi2016general,
  benkyi2019calculation} have applied the TD formalism in conjunction
with the HA. For rigid molecules, this is a viable alternative to
methods that perform on-the-fly dynamical propagations, as they often
require expensive electronic structure calculations to be done at
every time step and Hessians to be re-evaluated multiple
times~\cite{beguvsic2020fly, begusic2022Applic}. A popular and quite
successful approach is to use the propagator of a quantum harmonic
oscillator in the coordinate basis, the so-called Mehler kernel, to
evaluate the transition dipole correlation function analytically.
Though some studies~\cite{niu2010theory, baiardi2013general,
  cerezo2013harmonic} have derived an expression that incorporates
thermal effects (herein referred to as $\mu_T$),
others~\cite{tapavicza2016importance, benkyi2019calculation} have
neglected temperature dependence and derived it exclusively at
\SI{0}{\kelvin} ($\mu_{0}$). Though one might reasonably expect
$\mu_T$ to converge smoothly to $\mu_{0}$ when
$T \rightarrow \SI{0}{\kelvin}$, this is not the case since $\mu_T$
becomes ill-defined in this limit. $\mu_T$ is commonly expressed as a
fraction where the Mehler kernel is in the numerator and the partition
function is in the denominator and both approach $0$ as
$T \rightarrow \SI{0}{\kelvin}$, leading to an indeterminate form
$0/0$ and causes numerical instabilities at extremely low
temperatures. To circumvent the issue, \citet{de2018theoretical}
suggested to redefine certain terms in $\mu_T$ at temperatures lower
than \SI{10}{\kelvin} such that it would coincide with $\mu_{0}$.
Doing so, one effectively gets two separate expressions: one used at
very low temperatures (less than \SI{10}{\kelvin}) and another used
everywhere else. Although this solution is acceptable, because one
does not expect to see noticeable differences in spectra taken in the
\SIrange{0}{10}{\kelvin} 
interval, a deeper understanding of the cause of the issue as well as
potential consequences are highly desired.

Quantum dynamics of one-dimensional harmonic oscillators that uses the
Mehler kernel exhibits occasional time-discontinuities and sudden
phase jumps~\cite{horvathy2011maslov, thornber1998propagator,
  rosenfelder2021numerical}. This feature is well documented and
explained in many textbooks that introduce the path-integral
formalism~\cite{tannor2007introduction, kleinert2009path,
  schulman2012techniques}. Given that $\mu_T(t)$ corresponding to two
multidimensional harmonic oscillators is built from several
single-mode propagators, one expects these discontinuities to impact
vibronic spectra simulations. Thus, it seems natural to assess such
impact and develop a method that delivers continuous $\mu_T(t)$. While
one may argue that because such discontinuities do not arise in
quantum dynamics obtained by solving the time-dependent
Schr\"{o}dinger equation, that the use of the Mehler kernel ought to
be discarded; the propagator approach should be preserved as it allows
for easy consideration of thermal effects.

This paper is organized as follows. We begin by introducing key
equations used to simulate vibronic spectra within the time-dependent
approach. Next, we show how terms inside $\mu_T(t)$ can be factored
and canceled out under the HA to yield a well-behaved expression at
$T\rightarrow \SI{0}{\kelvin}$. For the sake of simplicity, we analyze
the one-dimensional model first before deriving an expression for the
thermal vibrational correlation function of a multidimensional
harmonic oscillator. After that, we look into the time-discontinuity
of $\mu_T(t)$ and propose a phase tracking scheme that effectively
amounts to the inclusion of a \emph{Maslov
  correction}~\cite{horvathy2011maslov} for all individual normal
modes. The result is a formula for thermal transition dipole
correlation function that is well-defined across all temperatures,
time-continuous, and numerically stable. It can be computed for
relatively large molecules while fully accounting for Duschinsky
rotations and displacements within the adiabatic Hessian model. We
illustrate our developments by simulating absorption spectra of
pentacene at \SI{4}{\kelvin} in the Franck--Condon approximation and
benzene at \SI{298}{\kelvin} in the Herzberg--Teller approximation and
compare them with their experimental counterparts. Atomic units are
used throughout the paper.

\section{Theory}
\label{sec:theory}

\subsection{Vibronic spectra in the time-domain}
\label{sec:vibr-spectra-time}

For the processes that are of interest herein, we apply the
semiclassical approach wherein light is described classically and
molecules quantum mechanically. The rates of transitions between
discrete states can be approximated from perturbation theory by using
the state-to-state form of Fermi's golden
rule~\cite{schatz2002quantum} and invoking the dipole approximation.
The resulting absorption cross section $\sigma_T(\omega)$ for
transitions from a set of initially occupied molecular states
$ \{\ket{\varphi_{\i}}\}$ at thermal equilibrium to a set of states
$\{\ket{\varphi_{\f}}\}$ can be given by the following
formula~\cite{cerezo2013harmonic, baiardi2016general,
  domcke2004conical, tannor2007introduction}
\begin{equation}
  \label{eq.FermiGolden}
  \sigma_{T}(\omega) = 
  \frac{4\pi^2 \omega}{3c}
  \sum_{\i,\f}P_{\text{i}}\abs{\mel{\varphi_{\i}}{\hat{\mu}_{M}}{\varphi_{\f}}}^{2}\delta(\omega-\omega_{\f\i}),
\end{equation}
where $c$ is the speed of light,
$P_{\i}(T) = {e^{-\frac{E_{\i}}{k_{\beta}T}}}/{Z(T)}$ is the Boltzmann
population of the initial state at temperature $T$, $k_{\beta}$ is the
Boltzmann constant, $Z(T)$ is the system's partition function,
$\hat{\mu}_{M}$ is the molecular dipole operator and
$\delta(\omega-\omega_{\f\i})$ is a Dirac delta function where
$\omega_{\f\i} = \left({E_{\f} - E_{\i}}\right)$ is the excitation
energy.

In the Born-Oppenheimer approximation, the molecular initial and final
quantum states are factorized as
$\phi_{\I}(\mathbf{r}|\mathbf{R})\chi_{\i}(\mathbf{R})$ and
$\phi_{\F}(\mathbf{r}|\mathbf{R})\chi_{\f}(\mathbf{R})$, respectively.
Here, $\mathbf{r}$ are the electronic and $\mathbf{R}$ are the nuclear
coordinates; capitalized subscripts denote electronic, while the
lowercase subscripts denote nuclear states. Because we are only
interested in the vibrational states, from hereon
$\chi_{\f}(\mathbf{R})$ corresponds with the vibrational part of the
nuclear state and $\mathbf{R}$ are nuclear internal degrees of
freedom.

The quantum transitions amplitudes
$\mel{\varphi_{\i}}{\hat{\mu}_{M}}{\varphi_{\f}}$ are the matrix
elements of the molecular dipole operator
\begin{equation}
  \hat{\mu}_{M}(\mathbf{r},\mathbf{R}) = \hat{\mu}_{e}(\mathbf{r}) + \hat{\mu}_{N}(\mathbf{R}),
\end{equation}
where $\hat{\mu}_{e}(\mathbf{r})$ and $\hat{\mu}_{N}(\mathbf{R})$ are
the electronic and nuclear dipole moment operators, respectively.

Assuming the Born-Oppenheimer factorization, the electronic degrees of
freedom can be integrated out to leave
\begin{align}
  \label{eq.rhoijgeneral}
  \abs{\mel{\phi_I\chi_{\i}}{\hat{\mu}_{M}}{\phi_F\chi_{\f}}_{\mathbf{rR}}}^{2}
  &= \abs{\mel{\chi_{\i}}{\vec{\gamma}_{\I\F}}{\chi_{\f}}_{\mathbf{R}}}^{2},
\end{align}
where
\begin{equation}
  \vec{\gamma}_{\I\F}(\mathbf{R})=\mel{\phi_{\I}}{\hat{\mu}_{e}}{\phi_{\F}}_{\mathbf{r}}
\end{equation} 
is the electronic transition dipole moment function and
$\langle \cdots \rangle_{\mathbf{R}}$ and
$\langle \cdots \rangle_{\mathbf{r}}$ denotes integration over
$\mathbf{R}$ and $\mathbf{r}$, respectively. Note that
$\hat{\mu}_{\text{N}}$ does not contribute due the orthogonality of
electronic state within the Born-Oppenheimer approximation. In what
follows, we suppress the subscripts in $\vec{\gamma}_{\I\F}$ assuming
the initial and final electronic state remain fixed.
 
The transition dipole moment function is a three dimensional vector
whose components are functions of nuclear geometry,
$\vec{\gamma}^{\intercal}(\mathbf{R}) = \left[ \gamma_{x}(\mathbf{R}),
  \gamma_{y}(\mathbf{R}), \gamma_{z}(\mathbf{R}) \right]$. Its
evaluation is simplified through the Condon approximation, where we
expand each component of $\vec{\gamma}(\mathbf{R})$ around a potential
energy minimum $\mathbf{R}_{0}$ in Taylor series and truncate to first
order,
\begin{equation}
  \label{eq:electransdipmom}
  \vec{\gamma}(\mathbf{R}) \approx \vec{\gamma}(\mathbf{R}_{0}) + \grad^{\intercal}\vec{\gamma}(\mathbf{R}_{0})(\mathbf{R}-\mathbf{R}_{0}),
\end{equation}
where
\begin{equation}
  \grad^{\intercal}\vec{\gamma}(\mathbf{R}_{0}) 
  =
  \begin{bmatrix}
    \frac{\partial\gamma_{x}(\mathbf{R})}{\partial{R}_{1}} &
    \frac{\partial\gamma_{x}(\mathbf{R})}{\partial{R}_{2}} & ... &
    \frac{\partial\gamma_{x}(\mathbf{R})}{\partial{R}_{N}}
    \\
    \frac{\partial\gamma_{y}(\mathbf{R})}{\partial{R}_{1}} &
    \frac{\partial\gamma_{y}(\mathbf{R})}{\partial{R}_{2}} & ... &
    \frac{\partial\gamma_{y}(\mathbf{R})}{\partial{R}_{N}}
    \\
    \frac{\partial\gamma_{z}(\mathbf{R})}{\partial{R}_{1}} &
    \frac{\partial\gamma_{z}(\mathbf{R})}{\partial{R}_{2}} & ... &
    \frac{\partial\gamma_{z}(\mathbf{R})}{\partial{R}_{N}}
    \\
  \end{bmatrix}
\end{equation}
and $N$ denotes the number of vibrational degrees of freedom.

Plugging this expansion into Eq.~(\ref{eq.rhoijgeneral}), one can
approximate
$\abs{\mel{\chi_{\i}}{\vec{\gamma}}{\chi_{\f}}_{\mathbf{R}}}^{2}$ as
the sums of three contributions
\begin{equation}
  \label{eq:decons_rho_if}
  \abs{\mel{\chi_{\i}}{\vec{\gamma}}{\chi_{\f}}_{\mathbf{R}}}^{2}
  = \rho_{\i\f}^{\FC} + \rho_{\i\f}^{\FCHT} + \rho_{\i\f}^{\HT},
\end{equation}
where
\begin{align}
  \label{eq:rhos_defns}
  & \rho_{\i\f}^{\FC} = \vec{\gamma}^{\intercal}(\mathbf{R}_{0})\cdot\vec{\gamma}(\mathbf{R}_{0})
    \abs{\braket{\chi_{\i}}{\chi_{\f}}_{\mathbf{R}}}^{2},\\
  &\rho_{\i\f}^{\FCHT} = 
    2\vec{\gamma}^{\intercal}(\mathbf{R}_{0})\grad^{\intercal}\vec{\gamma}(\mathbf{R}_{0})
    \braket{\chi_{\i}}{\chi_{\f}}_{\mathbf{R}}
    \mel{\chi_{\f}}{\mathbf{R}-\mathbf{R}_{0}}{\chi_{\i}}_{\mathbf{R}}, \\ 
  &\rho_{\i\f}^{\HT}  = 
    \mel{\chi_{\f}}{\mathbf{R}-\mathbf{R}_{0}}{\chi_{\i}}_{\mathbf{R}}^{\intercal}
    \grad\vec{\gamma}^{\intercal}(\mathbf{R}_{0})
    \grad^{\intercal}\vec{\gamma}(\mathbf{R}_{0})
    \mel{\chi_{\f}}{\mathbf{R}-\mathbf{R}_{0}}{\chi_{\i}}_{\mathbf{R}}
\end{align}
and the nuclear states $\chi(\mathbf{R})$ are real functions. The
superscripts FC, FCHT and HT in the previous equations refer to
Franck--Condon, Franck--Condon Herberg--Teller and Herzberg--Teller
terms, respectively.

By substituting Eq.~(\ref{eq:decons_rho_if}) and the Dirac delta
function as a Fourier transform
$\delta(\omega) = \frac{1}{2\pi} \int^{\infty}_{-\infty}e^{i\omega t}
dt$ into Eq.~(\ref{eq.FermiGolden}), the spectra is defined in the
time domain~\cite{peng2010vibration, niu2010theory,
  baiardi2013general}
\begin{equation}
  \sigma_T(\omega) = 
  \frac{2\pi \omega}{3c} 
  \Re \int_{0}^{\infty} e^{i\omega t} \left( \mu_T^{\FC}(t) +  \mu_T^{\FCHT}(t) + \mu_T^{\HT}(t) \right) dt,
\end{equation}
where
\begin{align}
  \label{eq:mus_defns_FC}
  & \mu_T^{\FC}(t) = \sum_{\i,\f}\frac{1}{Z_{\I}}\rho_{\i\f}^{\FC}e^{-iE_{\i}\tau}e^{{-iE_{\f}t}}, \\
  \label{eq:mus_defns_FCHT}
  & \mu_T^{\FCHT}(t) = \sum_{\i,\f}\frac{1}{Z_{\I}}\rho_{\i\f}^{\FCHT}e^{-iE_{\i}\tau}e^{{-iE_{\f}t}}, \\
  & \mu_T^{\HT}(t) = \sum_{\i,\f}\frac{1}{Z_{\I}}\rho_{\i\f}^{\HT}e^{-iE_{\i}\tau}e^{{-iE_{\f}t}}
    \label{eq:mus_defns_HT}
\end{align}
are hereon referred to as FC, FCHT and HT correlation functions,
respectively and $\tau=\frac{-i}{k_{\beta}T}-{t}$.

\subsection{$\bm{\mu_T}^{\bm{\FC}}(\bm{t})$ at low temperature}
\label{subsec:lowTemp_mu}

The analytical expression for $\mu_T^{\FC}(t)$ can be represented in
the continuous basis (see details in
Appendix~\ref{subsec:derivAnalymuFC_1}) as
\begin{equation}
  \label{eq:mu_FC_gen}
  {\mu}_T^{\FC}(t) 
  = 
  {\vec{\gamma}^{\intercal}(\mathbf{R}_{0})\cdot\vec{\gamma}(\mathbf{R}_{0})}
  \tilde{\mu}_T^{\FC}(t),
\end{equation}
where
\begin{equation}
  \label{eq:mu_FC_tilde_gen}
  \tilde{\mu}_T^{\FC}(t) 
  = \int\frac{1}{Z_{\I}}\mel{\mathbf{R'}}{e^{-i\hat{H}_{\I}\tau}}{\mathbf{R}}\mel{\mathbf{R}}{e^{-i\hat{H}_{\F}t}}{\mathbf{R'}}d\mathbf{R}d\mathbf{R'}.
\end{equation}

Before continuing with multidimensional cases, we turn our attention
to the generic one-dimensional harmonic oscillator.
We denote the spatial variable of the one-dimensional harmonic
oscillator as $q$, its frequency as $w$, its mass as $m$, and we can
evaluate its ${\tilde \mu}_T^{\FC}(t)$ as per
Eq.~(\ref{eq:mu_FC_tilde_gen}) using the partition function
\begin{align}
  Z_{\I} (\T) = 
  \frac{e^{-\frac{w}{2k_{\beta}T}}}{1-e^{-\frac{w}{k_{\beta}T}}}
  =\frac{1}{2\sinh
  \left(
  \frac{w}{2k_{\beta}T}
  \right)
  },
\end{align}
and the propagator in the position basis $\mel{q}{e^{-i\hat{H}t}}{q'}$
also known as the ``Mehler kernel," which has the well-known
form~\cite{schulman2012techniques}
\begin{equation}
  \label{eq:Mehler_kernel_1D}
  \mel{q}{e^{-i\hat{H}t}}{q'}= 
  \sqrt{\frac{m}{2\pi i}}\sqrt{\frac{w}{\sin\left(wt\right)}}e^{\left(imw\left(\frac{\left(q^2 + q'^2\right)\cot\left(wt\right)}{2}-\frac{qq'}{\sin\left(wt\right)}\right)\right)}.
\end{equation}

In the limit $T\rightarrow \SI{0}{\kelvin}$, both $Z_{\I}(\T)$ and
$\mel{q}{e^{-i\hat{H}\tau}}{q'}$ converge to 0. In the Mehler kernel,
this behavior is due to its pre-exponential factor
\begin{align}
  \sqrt{\frac{1}{2i\sin\left(w\tau\right)}} = \frac{e^{\frac{iwt}{2}}e^{-\frac{w}{2k_{\beta}T}}}{\sqrt{1-e^{-2iw\tau}}}.
\end{align}

As written,
$\lim_{T\rightarrow 0} \frac{\mel{q}{e^{-i\hat{H}t}}{q'}}{Z(T)}$ has
an indeterminate form $0/0$. This is not only a conceptual problem,
but it also makes computations susceptible to numerical instabilities
at extremely low temperatures.

To circumvent this issue, some studies (\latin{e.g.}
Ref.~\citenum{de2018theoretical}) have employed two different
expressions for ${\mu}_T^{\FC}(t)$. One is intended for high
temperatures where one does not encounter the aforementioned numerical
issues and coincides with the derivations presented so far. Another is
meant for temperatures near \SI{0}{\kelvin} and is derived from
Eq.~({\ref{eq.FermiGolden}}) but with the sum extending only through
the vibrational states of the final electronic state, which
effectively assumes that the initial state is entirely populated by
the ground vibrational state.

We propose a well-defined expression for ${\mu}_T^{\FC}(t)$, namely
by evaluating the determinate form of the quotient
$\widetilde{K}_{\I}(q,q',\tau)$,
\begin{align}
  \label{eq:Ktilde_1D}
  \widetilde{K}_{\I}(q,q',\tau) 
  &= \frac{\mel{q}{e^{-i\hat{H_{\I}}\tau}}{q'}}{Z_{\I}} \\
  \label{eq:Ktilde_1D_full}
  &=\sqrt{\frac{w_{\I}}{\pi}}\frac{\left(1-e^{-\frac{w_{\I}}{k_{\beta} T}}\right)e^{\frac{iw_{\I}t}{2}}}{\sqrt{1-e^{-2iw_{\I}\tau}}}
    e^{\left(iw_{\I}\left(\frac{\left(q^2 + q'^2\right)cot\left(w_{\I}\tau\right)}{2}-\frac{qq'}{\sin\left(w_{\I}\tau\right)}\right)\right)} 
\end{align}
in which the term $e^{-\frac{w}{2k_{\beta}T}}$ is canceled out in both
$Z_{\I}(\T)$ and $\mel{q}{e^{-i\hat{H}\tau}}{q'}$. Substituting
Eq.~(\ref{eq:Ktilde_1D_full}) into Eq.~(\ref{eq:mu_FC_tilde_gen})
yields an expression of the FC correlation function that is
well-defined at low-temperatures
\begin{equation}
  \label{eq:muFCbar}
  \bar{\mu}_T^{\FC}(t)= 
  {\vec{\gamma}^{\intercal}({q}_{0})\cdot\vec{\gamma}({q}_{0})}
  \int \widetilde{K}_{\I}(q',q,\tau) \mel{q}{e^{-i\hat{H_\mathrm{F}}t}}{q'} dqdq'.
\end{equation}

It is worth noting that one can also \textit{partially} derive
Eq.~(\ref{eq:Ktilde_1D_full}) by neglecting or removing the zero-point
energy in the Mehler kernel and in $Z_{\I}(\T)$. We can do so by
multiplying both expressions by $e^{{iE_{0}^{\text{HO}}t}}$ where
$E_{0}^{\text{HO}}$ is the zero-point energy of the harmonic
oscillator,
\begin{align}
  \label{eq:KnoZPE}
  K_{\noZPE}(q,q',t)
  & =  
    e^{{iE_{0}^{\text{HO}}t}} 
    \mel{q}{e^{-i\hat{H}t}}{q'}  \\
  & = 
    \sqrt{\frac{mw}{\pi}}\frac{1}{\sqrt{1-e^{-2i wt}}}e^{\left(imw\left(\frac{\cot(wt)}{2}\left(q^2+q'^2\right)-\frac{1}{\sin(wt)}qq'\right)\right)},
    \label{eq:KnoZPE_HO}
  \\
  \label{eq:ZnoZPE}
  Z_{\noZPE}(\T) &= e^{\frac{{E_{0}^{\text{HO}}}}{k_{\beta}T}} Z(\T)  \\
  &= \frac{1}{1-e^{-\frac{w}{k_{\beta}T}}},
    \label{eq:ZnoZPE_HO}
\end{align}
so that
\begin{equation}
  \label{eq:general_Cancel}
  \widetilde{K}(q,q',\tau) = e^{{iE_{0}t}}\frac{K_{\noZPE}(q,q',\tau)}{Z_{\noZPE}(\T)},
\end{equation}
thereby making the quotient of $K_{\noZPE}$ and $Z_{\noZPE}$ differs
from $\widetilde{K}$ by a global phase. This phase does not alter the
spectral line shape, it merely shifts it horizontally by
$E_{0}^{\text{HO}}$ along the energy axis.

The issue of the indeterminate form of
$\lim_{\T\rightarrow 0} \frac{\mel{q}{e^{-i\hat{H}\tau}}{q'}}{Z(\T)}$
is general as it is not limited to the harmonic oscillator model and
arises in all bounded systems. The generic forms of
$\mel{q}{e^{-i\hat{H}t}}{q'}$ and $Z(T)$ for any bounded system can be
expressed in the eigenbasis $\{\phi_{n}(q)\}$ and energy spectrum
$\{E_{n}\}$ of the Hamiltonian
\begin{align}
  \mel{q}{e^{-i\hat{H}\tau}}{q'} 
  &= 
    e^{-\frac{E_{0}}{k_{\beta}T}}e^{-iE_{0}t}
    \sum_{n}e^{i \Delta E_{n}\tau}\phi_{n}^{*}(q)\phi_{n}(q'), \\
  Z(\T)  
  &=
    e^{-\frac{E_{0}}{k_{\beta}T}} 
    \sum_{n} e^{-\frac{\Delta E_{n}}{k_{\beta}T}},
\end{align}
where $E_{0}$ is the system's zero point energy, $\Delta E_{n}$ is the
energy difference between $E_{n}$ and $E_{0}$ and
$\lim_{\T\rightarrow 0} e^{-\frac{E_{0}}{k_{\beta}T}}\rightarrow 0$.
As a result, Eq.~(\ref{eq:KnoZPE}), Eq.~(\ref{eq:ZnoZPE}) and
Eq.~(\ref{eq:general_Cancel}) are general equations in the sense that
they are valid to all bounded systems --- one only needs to replace
$E_{0}^{\text{HO}}$ with $E_{0}$. These equations used in conjunction
with Eq.~(\ref{eq:muFCbar}) can effectively yield Franck-Condon
correlation functions with a determinate forms. Moreover, it is
evident that when partition functions and propagators are obtained
approximately, that the cancellation in Eq.~(\ref{eq:general_Cancel})
only works if both entail the exact same zero-point energy.

\subsection{$\bm{\tilde{\mu}_T}^{\bm{\FC}}(\bm{t})$ in the
  multidimensional case}
\label{subsec:mu_multi_Dim}

Eq.~(\ref{eq:Mehler_kernel_1D}), may be re-expressed more succinctly
in mass-weighted coordinates as
\begin{equation}
  K(q,q',t) = \sqrt{\frac{a}{2\pi i}}e^{\left(\frac{i}{2}
      \begin{bmatrix}
	q & q' \\
      \end{bmatrix}
      \begin{bmatrix}
	b & -a \\
	-a & b
      \end{bmatrix}
      \begin{bmatrix}
	q \\
	q'
      \end{bmatrix}
    \right)  
  }
\end{equation}
where $a=\frac{w}{\sin(wt)}$ and $b=\frac{w}{\tan(wt)}$. In the case
of a $N$-dimensional harmonic oscillator, if one works in the basis of
normal modes $\mathbf{Q}$, the Hamiltonian can be expressed as the sum
of $N$ one-dimensional Hamiltonians. The corresponding $N$-dimensional
propagator is thus the product of one-dimensional ones,
\begin{equation}
  \label{eq:KQQ} 
  \begin{aligned} 
    K(\mathbf{Q},\mathbf{Q'},t) &= \prod_{n=1}^{N}K_{n}(q_n, q_{n}',t) \\
    &= \sqrt{\frac{\det(\mathbf{A})}{\left(2\pi
          i\right)^{N}}}e^{\left(\frac{i}{2}
	\begin{bmatrix}
          \mathbf{Q} & \mathbf{Q'} \\
	\end{bmatrix}
	\begin{bmatrix}
          \mathbf{B} & -\mathbf{A} \\
          -\mathbf{A} & \mathbf{B}
	\end{bmatrix}
	\begin{bmatrix}
          \mathbf{Q} \\
          \mathbf{Q'}
	\end{bmatrix}
      \right) }
  \end{aligned}
\end{equation} 
where $\mathbf{A}$ and $\mathbf{B}$ are diagonal matrices with
elements $A_{n}=\frac{w_{n}}{\sin(w_{n}t)}$ and
$B_{n}=\frac{w_{n}}{\tan(w_{n}t)}$, respectively. In the case of
$\widetilde{K}(\mathbf{Q},\mathbf{Q}',\tau)$, one obtains a similar
expression but with a different pre-exponential factor,
\begin{equation}
  \label{eq:KtildeQQ}
  \widetilde{K}(\mathbf{Q}, \mathbf{Q}', t) = \det(\mathbf{\Omega})e^{\left(\frac{i}{2}
      \begin{bmatrix}
	\mathbf{Q} & \mathbf{Q'} \\
      \end{bmatrix}
      \begin{bmatrix}
	\mathbf{B} & -\mathbf{A} \\
	-\mathbf{A} & \mathbf{B}
      \end{bmatrix}
      \begin{bmatrix}
	\mathbf{Q} \\
	\mathbf{Q'}
      \end{bmatrix}
    \right) 
  }
\end{equation}
where $\mathbf{\Omega}$ is a diagonal matrix with elements
${\Omega}_{n}=\sqrt{\frac{w_{n}}{\pi}}\frac{\left(1-e^{-w_{n}\beta}\right)e^{\frac{iw_{n}t}{2}}}{\sqrt{1-e^{2iw_{n}t}e^{-2w_{n}\beta}}}$.

For molecules under the HA, nuclear degrees of freedom are often
represented in the basis of normal modes. Those coincide with the
eigenvectors of mass-weighted Hessians of either initial or final
electronic states, which are denoted by $\mathbf{\Lambda}_{\I}$ and
$\mathbf{\Lambda}_{\F}$, respectively. Given an initial nuclear
configuration expressed in the basis of normal modes of the final
electronic state $\mathbf{Q}_{\F}$, a change in basis to the normal
modes of the initial electronic state $\mathbf{Q}_{\I}$ can be done
via the Duschinsky transformation
\begin{equation}
  \label{eq:DuschinskyRotQI}
  \mathbf{Q}_{\I} = \mathbf{J}_{\I\F}\mathbf{Q}_{\F}+\mathbf{K}
\end{equation} 
where
$\mathbf{J}_{\I\F} =
\mathbf{\Lambda}_{\I}^{\intercal}\mathbf{\Lambda}_{\F}$ and
$\mathbf{K} = \mathbf{\Lambda}_{\I}^{\intercal}\mathbf{L_{\F}}$ with
$\mathbf{L}_{\F}$ being the nuclear displacement of the final state
PES minimum expressed in mass-weighted Cartesian coordinates.

Ultimately, by solving the integral resulting from the insertion of
Eq.~(\ref{eq:KQQ}) and Eq.~(\ref{eq:KtildeQQ}) into multidimensional
analogue of Eq.~(\ref{eq:muFCbar}) and the change of variables as per
Eq.~(\ref{eq:DuschinskyRotQI}), we get
\begin{equation}
  \label{eq:muFC_final}
  \begin{aligned}
    \bar{\mu}_T^{\FC}(t) &= \det(\mathbf{\Omega}_{\I})
    \sqrt{\frac{(-2\pi i)^{N}\det(\mathbf{A}_{\F})}{\det(\mathbf{D})}}
    e^{\left(i\left(
          -\frac{\mathbf{E}^{\intercal}\mathbf{D}^{-1}\mathbf{E}}{2} +
          {F} \right)\right)}
  \end{aligned}
\end{equation} 
where
\begin{equation}
  \label{eq:muFCcomponents}
  \begin{aligned}
    \mathbf{D} &=
    \begin{bmatrix}
      \mathbf{B}_F + \mathbf{J}_{\I\F}^{\intercal}\mathbf{B}_I\mathbf{J}_{\I\F} & -(\mathbf{A}_F + \mathbf{J}_{\I\F}^{\intercal}\mathbf{A}_I\mathbf{J}_{\I\F}) \\
      -(\mathbf{A}_F + \mathbf{J}_{\I\F}^{\intercal}\mathbf{A}_I\mathbf{J}_{\I\F}) & \mathbf{B}_F + \mathbf{J}_{\I\F}^{\intercal}\mathbf{B}_I\mathbf{J}_{\I\F} \\
    \end{bmatrix}, \\
    \mathbf{E}^{\intercal} &=
    \begin{bmatrix}
      \mathbf{K}^{\intercal}(\mathbf{B}_{\I}-\mathbf{A}_{\I})\mathbf{J}_{\I\F} & \mathbf{K'}^{\intercal}(\mathbf{B}_{\I}-\mathbf{A}_{\I})\mathbf{J}_{\I\F} \\
    \end{bmatrix}, \\
    {F} &= \mathbf{K'}^{\intercal}
    (\mathbf{B}_{\I}-\mathbf{A}_{\I})\mathbf{K}
  \end{aligned}
\end{equation} 
For large molecules, the determinants in Eq.~(\ref{eq:muFC_final}) may
acquire large magnitudes and their computations may result in overflow
errors. Thus, to compute $\mu_T^{\FC}(t)$ of large molecules, it is
necessary to reexpress the pre-exponential factor in a more tractable
form. Looking at Eq.~(\ref{eq:muFC_final}), one may be tempted to
compute $\det(\mathbf{\Omega}_{\I}^{2}\mathbf{A}_{\F}\mathbf{D}^{-1})$
and expect a numerical outcome whose magnitude is close to $1$, given
that the matrix elements of $\mathbf{\Omega}_{\I}^{2}\mathbf{A}_{\F}$
are of approximately the same order of magnitude as those of
$\mathbf{D}$. However, this cannot be done because $\mathbf{D}$ is a
$2N \times 2N$ matrix and $\mathbf{\Omega}_{\I}$ and $\mathbf{A}_{\F}$
are $N \times N$ matrices.
 
To proceed forward approximately along this path, we need to define an
$N \times N$ matrix whose determinant equals that of $\mathbf{D}$.
Such matrix is
$\widetilde{\mathbf{D}} = \mathbf{D_{11}} \left(\mathbf{D_{11}} -
  \mathbf{D_{12}}\mathbf{D_{11}}^{-1}\mathbf{D_{12}}\right)$, which is
constructed from the submatrices of $\mathbf{D}$ where
$\mathbf{D_{11}}= \left(\mathbf{B}_F +
  \mathbf{J}_{\I\F}^{\intercal}\mathbf{B}_I\mathbf{J}_{\I\F}\right)$,
$\mathbf{D_{12}} = -(\mathbf{A}_F +
\mathbf{J}_{\I\F}^{\intercal}\mathbf{A}_I\mathbf{J}_{\I\F})$ and
satisfies $\det( \mathbf{D} ) = \det( \widetilde{\mathbf{D}} )$ and
$(\det(\mathbf{D}))^{-1} = \det( \widetilde{\mathbf{D}}^{-1}
)$~\cite{de2018theoretical, powell2011calculating}. Furthermore, to
ensure that the use of $\sqrt{\det(\mathbf{\Omega}_{\I}^{2})}$ gives
the same result as that of $\det{\mathbf{\Omega}_{\I}}$, we introduce
$\eta$ that equals either $+1$ or $-1$ such that
$\det(\mathbf{\Omega}_{\I}) = \eta
\sqrt{\det(\mathbf{\Omega}_{\I}^{2})}$. The substitution of these
terms inside Eq.~(\ref{eq:muFC_final}) yields
\begin{equation}
  \label{eq:muFCfinal2}
  \begin{aligned}
    \bar{\mu}_T^{\FC}(t) &= \eta (-2\pi i )^{\frac{N}{2}} \sqrt{
      \det( \mathbf{\Omega}_{\I}^{2} \mathbf{A}_{\F}
      \widetilde{\mathbf{D}}^{-1} ) } e^{\left(i\left(
          -\frac{\mathbf{E}^{\intercal}\mathbf{D}^{-1}\mathbf{E}}{2} +
          {F} \right)\right)},
  \end{aligned}
\end{equation}
which is a numerically more stable than Eq.~(\ref{eq:muFC_final}).

\subsection{Time-discontinuity in
  $\bm{\tilde{\mu}_T}^{\bm{\FC}}(\bm{t})$}
\label{subsec:TdisContinMuFC}

A direct numerical implementation of Eq.~(\ref{eq:muFCfinal2}) yields
a discontinuous $\bar{\mu}_T^{\FC}(t)$ and an erroneous vibronic
spectrum. The source of the time-discontinuity is the pre-exponential
factor. The term under the square root is a complex-valued function of
time, $|S(t)|e^{i\theta(t)}$ and has two square roots:
$\sqrt{|S(t)|}e^{i\frac{\theta(t)}{2}}$ and
$\sqrt{|S(t)|}e^{i\left(\frac{\theta(t)}{2}+\pi\right)}$. When a
square root is computed, most numerical libraries output the principal
value, which is usually the root with the positive real component.
Such choice is not always valid.

Fundamentally, the discontinuity arises because the expression used
for the one-dimensional harmonic oscillator propagator,
Eq.~(\ref{eq:Mehler_kernel_1D}), can be discontinuous in time; it is
multiplied by $\frac{1}{\sqrt{-1}}=e^{-i\pi/2}$ after half-period
increments~\cite{thornber1998propagator}. In the path integral (PI)
formalism, these extra phase jumps stem from the contributions of
variations surrounding classical trajectories; see
Appendix~\ref{subsec:time-discon-PI} for the detailed explanation.

To correct for the discontinuity in the one-dimensional harmonic
oscillator propagator, it is common to employ the \emph{Maslov phase
  correction}~\cite{horvathy2011maslov}. This correction consists in
adopting the following expression for the Mehler kernel instead of
Eq.~(\ref{eq:Mehler_kernel_1D}):
\begin{align}
  \label{eq:Maslov_Correc} 
  \mel{q}{e^{-i\hat{H}t}}{q'} = \sqrt{\frac{m}{2\pi i}}\sqrt{\frac{w}{\abs{\sin\left(wt\right)}}}e^{\left(imw\left(\frac{\left(q^2 + q'^2\right)\cot\left(wt\right)}{2}-\frac{qq'}{\abs{\sin\left(wt\right)}}\right)\right)}e^{\frac{-i\nu\pi}{2}},
\end{align}
where $\nu$ is called the \emph{Morse index} and it is the number of
elapsed half-periods in the time span
$\left[0,t\right]$~\cite{tannor2007introduction, kleinert2009path,
  schulman2012techniques}. In deriving $\mu_T^{\FC}(t)$ of a
multidimensional harmonic oscillator, one could have begun with
Eq.~(\ref{eq:Maslov_Correc}) rather than
Eq.~(\ref{eq:Mehler_kernel_1D}). This would have produced an
expression analogous to Eq.~\eqref{eq:muFCfinal2} that would have
included $e^{\frac{-i\nu\pi}{2}}$ terms for every initial and final
electronic state normal modes. Such an approach would have had the
advantage of preventing the time-discontinuity issue. The
disadvantage, however, is that one would have had to keep track of all
Morse indexes and evolve them over the course of the dynamics, thereby
making the entire procedure cumbersome and error prone. Instead, we
propose an alternative approach of tracking the \emph{total}
phase. 
We multiply the right-hand side of Eq.~(\ref{eq:muFCfinal2}) with a
real phase factor $\theta(t)$. This phase factor will equal either $1$
or $-1$, depending on which value minimizes the difference between
$\mu_T^{\FC}(t)$ values at consecutive time steps. This effectively
allows one to select the appropriate branch of $z^{\frac{1}{2}}$ and
to ensure the continuity of real and imaginary parts of
$\mu_T^{\FC}(t)$, which, as we will show in
Sec.~\ref{subsec:pentacene}, is paramount to vibronic spectra
simulations. The inclusion of the phase factor yields a final
expression for the FC correlation function
\begin{equation}
  \label{eq:muFCfinal3}
  \tilde{\mu}_T^{\FC}(t)
  =
  \theta
  \eta (-2\pi i )^{\frac{N}{2}}
  \sqrt{
    \det(
    \mathbf{\Omega}_{\I}^{2}
    \mathbf{A}_{\F}
    \widetilde{\mathbf{D}}^{-1}
    )
  }
  e^{\left( i\left(-\frac{\mathbf{E}^{\intercal}\mathbf{D}^{-1}\mathbf{E}}{2} + {F} \right)\right)},
\end{equation}
which is applicable for any temperature $T \ge \SI{0}{\kelvin}$, is
continuous in time, and is stable numerically for large molecules with
many vibrational degrees of freedom. The phase tracking procedure
removes the need to compute the Morse indexes of all normal modes at
all time steps. However, assessing the continuity between
$\mu_T^{\FC}(t)$ values of adjacent time steps requires that dynamical
simulations be performed with sufficiently small time steps.

In the case where several normal modes have the same frequencies,
degenerate Hessian eigenvectors are arbitrarily defined. This
arbitrariness can find itself in the Duschinsky matrix
$\mathbf{J}_{\I\F}$ of Eq.~(\ref{eq:DuschinskyRotQI}), and in
$\tilde{\mu}_T^{\FC}(t)$ of Eq.~(\ref{eq:muFCfinal3}). The
degeneracies could cause $\tilde{\mu}_T^{\FC}(t)$ to have an
arbitrary phase, which would be unresolved by the phase-tracking
procedure herein. In our implementation, we avoid this issue
altogether by computing the normal modes once at the start of the
dynamics. The Duschinsky matrix is computed once, and it is stored in
memory and reused in subsequent time steps. This effectively amounts
to fixing the basis of the degenerate subspace.

\section{Results and Discussion}
\label{sec:results-discussion}

\subsection{Computational details}
\label{sec:comp-deta}

We simulated the absorption spectra of pentacene and benzene.
Electronic structure calculations were performed using
Firefly~\cite{FFly} software package. Ground and excited state
Hessians and geometries minima were computed at
B3LYP/6-311G(df)~\cite{becke1993new, lee1988development,
  ditchfield1971self} level of theory. This level of theory is
comparable to ones used in other theoretical simulations of vibronic
spectral shapes that report good agreement with their experimental
counterparts~\cite{de2018theoretical, cerezo2013harmonic,
  cerezo2016revisiting}. Additionally, it was
demonstrated~\cite{benkyi2019calculation} that this level of theory
produces spectral line shapes similar to those obtained with the
second-order approximate coupled-cluster (CC2) method and hybrid
functionals used in conjunction with larger basis
sets~\cite{jacquemin2011td}.

Hessians were computed by
first-order numerical differentiation of gradients and atomic
coordinate displacements of \SI{0.01}{\angstrom}. For each Hessian
calculation, a total of 3N+1 gradients were evaluated (N here is the
number of atoms in the molecule). Excited state calculations were
performed with TDDFT in which the first 10 roots were computed. For
each molecule, the lowest singlet excited state was selected as it
corresponded to the state of interest.

The total duration of dynamical simulations were
\SI{2000}{\femto\second} with a time step of \SI{0.005}{\fs}. This
time step is chosen for the phase tracking procedure to avoid any
discontinuities; it is comparable in magnitude to ones used in
previous studies; \SI{0.006}{\fs} in Ref.~\citenum{baiardi2013general}
and \SI{0.015}{\fs} in Ref.~\citenum{de2018theoretical}. To model
broadening effects stemming from environmental factors present in
experiments, $\mu_T^{\FC}(t)$, $\mu_T^{\FCHT}(t)$ and
$\mu_T^{\HT}(t)$ are multiplied with an exponential damping
function, $f(t) = e^{-\frac{t}{\kappa}}$, where $\kappa$ is the
relaxation time. For pentacene and benzene respectively, we used
$\kappa = \SI{138.2}{\fs}$ and $\kappa = \SI{40.3}{\fs}$. The discrete
Fourier transform was performed using the NumPy suite of numerical
libraries~\cite{2020NumPy-Array}.

\subsection{Absorption spectra of pentacene in the Franck--Condon
  approximation}
\label{subsec:pentacene}

\begin{figure}
  \includegraphics[scale=0.7]{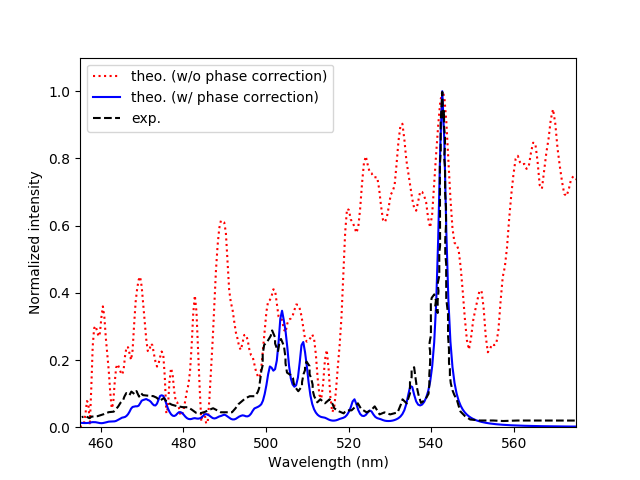}
  \caption{Absorption spectra for pentacene at $\SI{4}{\kelvin}$. The
    red dotted line corresponds to a theoretical simulation that does
    not incorporate phase correction and the solid blue line includes
    the phase correction. The theoretical spectrum with phase
    correction was shifted horizontally by $\SI{0.56}{eV}$ so as to
    match the experimental spectra~\cite{thusek2021high}.}
  \label{fig:pentacene_ab_theo_exp_4K}
\end{figure}

The vibronic spectra of pentacene has been the subject of numerous
studies~\cite{halasinski2000electronic, niu2010theory,
  benkyi2019calculation, thusek2021high} and to demonstrate the
applicability of Eq.~(\ref{eq:muFCfinal3}) in low temperatures, we
compare in Fig.~\ref{fig:pentacene_ab_theo_exp_4K} the theoretical
spectra of pentance at \SI{4}{\kelvin} obtained with and without phase
correction to its experimental counterpart~\cite{thusek2021high}.
Pentacene belongs to the $D_{2h}$ symmetry group and the featured
spectra correspond to a $ A_{g} \rightarrow B_{2u}$ transition that is
Franck--Condon allowed. The experimental spectrum has been obtained
using the matrix-isolation technique where pentacene monomers were
trapped in a neon matrix. At low concentrations, this methodology
approximates light absorption and emission in the absence of solvation
and environmental effects. Thus, DFT and TDDFT calculations were done
in vacuum. The phase-corrected theoretical spectrum was shifted by
\SI{0.56}{\electronvolt} 
as to improve correspondence with experiment. The need to shift
theoretical spectra stems from the general tendency of B3LYP to
overestimating singlet vertical excitations of benzene and acenes by
approximately \SI{0.5}{\electronvolt}~\cite{adamo1999accurate,
  benkyi2019calculation}.

As clearly visible in Fig.~\ref{fig:pentacene_ab_theo_exp_4K}, absence
of phase correction results in a nonsensical spectrum (the red dotted
curve). The solid blue line however includes the phase correction for
which the branch has been appropriately chosen to render the
correlation function continuous in time. These results indicate that
the continuity of real and imaginary components of the simulated
correlation function has a dramatic impact on vibronic spectra and
that the methodology based on Eq.~(\ref{eq:muFCfinal3}) is in
agreement with experiments.

\subsection{Absorption spectra of benzene with Herzberg--Teller
  effects}
\label{sec:absorpt-spectra-benz}

\begin{figure}
  \includegraphics[scale=0.7]{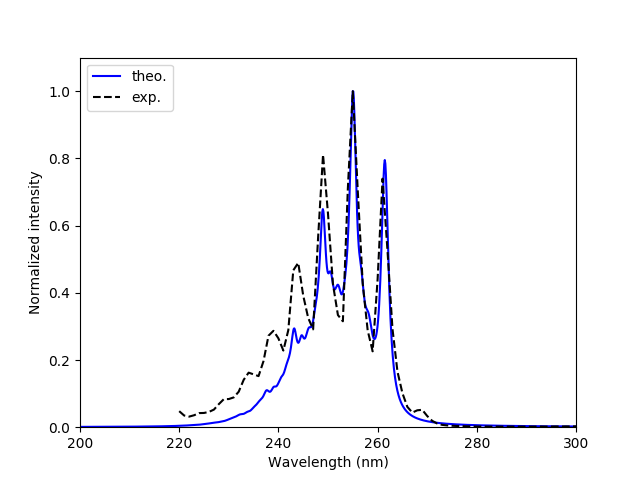}
  \caption{Absorption spectra for benzene in hexane at
    $\SI{298}{\kelvin}$. The solid blue line corresponds to a
    theoretical simulation that includes the phase correction. It was
    shifted horizontally by $\SI{0.53}{eV}$ so as to match the
    experimental spectra~\cite{taniguchi2018photochemcad}.}
  \label{fig:benzene_ab_theo_exp_300K}
\end{figure}

Benzene and its derivatives are the building blocks of a wide range of
organic molecules and have been the subject of numerous
investigations~\cite{de2018theoretical, berger1998calculation,
  bernhardsson2000theoretical, schumm2000franck, li2010symmetry,
  he2011franck}. Since benzene has $D_{6h}$ symmetry, its lowest
energy transition $A_{1g} \rightarrow B_{2u}$ is symmetry-forbidden in
the Franck--Condon approximation and the spectra arises solely from
the HT correlation function (see Eq.~(\ref{eq:muHTinterm-1}) of
Appendix~\ref{subsec:derivAnalymuHT}). We present in
Fig.~\ref{fig:benzene_ab_theo_exp_300K} the absorption spectra of
benzene at \SI{298}{\kelvin} for the purpose of illustrating the
application of Eq.~(\ref{eq:muFCfinal3}) (which is necessary to
compute $\mu_T^{\HT}(t)$ as shown in
Appendix~\ref{subsec:derivAnalymuHT}) at room temperature. In
Fig.~\ref{fig:benzene_ab_theo_exp_300K}, the theoretical spectrum was
shifted by \SI{0.53}{\electronvolt} so as to match the
experiment~\cite{taniguchi2018photochemcad}.
Our results seem to be in reasonable agreement with their experimental
counterpart as well as previous theoretical
reports~\cite{de2018theoretical, huh2017cumulant}.

\section{Conclusions}
\label{sec:conclusions}

We investigated two complications that may arise while computing
thermal vibrational correlation functions in the harmonic
approximation. The first complication is that the correlation
functions can be ill-defined at extremely low temperatures if the
partition function and the Mehler kernel, which are their key
components, are considered independently. This is caused by zero-point
energy terms present in both the partition function and the
propagator. Their removal, either through a cancellation of terms or
by neglecting the zero-point energy entirely, yields a thermal
correlation function that has a determinate form in the
$T\rightarrow \SI{0}{\kelvin}$ limit. This provides a means of
computing vibronic spectra within the time-dependent formulation that
is unified without resorting to consider two disjoint temperature
regimes.

The second complication is that thermal vibrational correlation
functions in molecular systems can be discontinuous in time. This
occurs because the pre-exponential factor of the Mehler kernel is
multivalued. Though one may attempt to construct correlation functions
from Maslov-corrected single mode propagators, this would necessitate
evolving all underlying Morse indexes during dynamics, which may be
cumbersome to implement. Alternatively, as we have shown, the
inclusion of a single \emph{global} phase parameter in the final
expression of the correlation function suffices to produce a result
whose real and imaginary parts are continuous. This phase ultimately
selects the appropriate branch of $z^{\frac{1}{2}}$ and minimizes the
difference between values of the correlation function of adjacent time
steps. While the current work was under review,
\citet{begusic2022Applic} published a study that also addressed in
part the discontinuity of the correlation function.

By addressing these two complications, we obtained an expression for
the thermal dipole autocorrelation function,
Eq.~\eqref{eq:muFCfinal3}, that is well-defined and continuous in
time. We assessed our findings on two realistic molecular systems,
pentacene at \SI{4}{\kelvin} and benzene at \SI{298}{\kelvin}. We
found that nonsensical absorption spectra were obtained when phase
corrections were omitted (as in the case of pentacene), while with the
phase correction our theoretical results matched well with the
experimental counterparts.

Although our work focused mostly on harmonic systems, our conclusions
are relevant to other models. In any model with positive zero-point
energy, one should expect that independent considerations of
propagators and partition functions can lead to thermal correlation
functions with indeterminate forms in the
$T \rightarrow \SI{0}{\kelvin}$ limit. In such case, one can resort to
the procedure described in Sec.~\ref{subsec:lowTemp_mu} to obtain a
determinate form. We also expect the time-discontinuity issue to
persist in bound anharmonic models whose potential energy surfaces are
locally well approximated by quadratic functions and where the
underlying dynamics are confined to those regions. This would be the
case if the initial and final potential energy surfaces resemble Morse
potentials with overlapping minima and large dissociation energies.
Furthermore, as it may be difficult to predict analytically the exact
instance of the propagator's sudden phase change in arbitrary
potentials, the numerical procedure discussed in
Sec.~\ref{subsec:TdisContinMuFC} is a general solution for
obtaining continuous correlation functions.


\appendix

\section{Derivation of the analytical expression of
  $\tilde{\bm{\mu}_T}^{\FC}(t)$}
\label{subsec:derivAnalymuFC_1}

By plugging the expression of $\rho_{\i\f}^{\FC}$ from
Eq.~(\ref{eq:rhos_defns}) inside that of $\mu_T^{\FC}(t)$ in
Eq.~(\ref{eq:mus_defns_FC}) we get
\begin{equation}
  \mu_T^{\FC}(t) = \vec{\gamma}^{\intercal}(\mathbf{R}_{0})\cdot\vec{\gamma}(\mathbf{R}_{0})\cdot\tilde{\mu}_T^{\FC}(t),
\end{equation}
where $\tilde{\mu}_T^{\FC}(t)$ is defined as
\begin{equation}
  \tilde{\mu}_T^{\FC}(t) = \sum_{\i,\f} \frac{1}{Z_{\I}}\bra{\chi_{\i}}\ket{\chi_{\f}}\bra{\chi_{\f}}\ket{\chi_{\i}}e^{-iE_{\i}\tau}e^{-iE_{\f}t}.
\end{equation}

By considering that
\begin{equation}
  \label{eq:AvgPropH_FChi_i}
  \sum_{\f} e^{-iE_{\f}t}\bra{\chi_{\i}}\ket{\chi_{\f}}\bra{\chi_{\f}}\ket{\chi_{\i}}
  =
  \mel{\chi_{\i}}{e^{-i\hat{H}_{\F}t}}{\chi_{\i}},
\end{equation}
we can rewrite $\tilde{\mu}_T^{\FC}(t)$ as
\begin{equation}
  \begin{aligned}
    \tilde{\mu}_T^{\FC}(t) = \frac{1}{Z_{\I}}
    \sum_{\i}e^{-iE_{\i}\tau}\mel{\chi_{\i}}{e^{-i\hat{H}_{\F}t}}{\chi_{\i}},
  \end{aligned}
\end{equation}
and through a resolution of the identity
$\hat{\mathbb{1}}=\int \ket{\mathbf{Q}}\bra{\mathbf{Q}} d\mathbf{Q}$
inside the equation above, we get
\begin{equation}
  \tilde{\mu}_T^{\FC}(t) = \frac{1}{Z_{\I}}\int \sum_{\i}e^{-iE_{\i}\tau}\bra{\chi_{\i}}\ket{\mathbf{Q}}\mel{\mathbf{Q}}{e^{-i\hat{H}_{\F}t}}{\mathbf{Q'}}\bra{\mathbf{Q'}}\ket{\chi_{\i}} d\mathbf{Q}d\mathbf{Q'}
\end{equation}
whose infinite sum via the spectral representation of
$e^{-i\hat{H}_{\i}t}=\sum_{\i}e^{-iE_{\i}t}\ket{\chi_{\i}}\bra{\chi_{\i}}$,

\begin{equation}
  \tilde{\mu}_T^{\FC}(t) = \frac{1}{Z_{\I}}\int\mel{\mathbf{Q'}}{e^{-i\hat{H}_{\I}\tau}}{\mathbf{Q}}\mel{\mathbf{Q}}{e^{-i\hat{H}_{\F}t}}{\mathbf{Q'}}d\mathbf{Q}d\mathbf{Q'}.
\end{equation}

\section{Derivation of the analytical expression of
  ${\bm{\mu_T}}^{\FCHT}(t)$}
\label{subsec:derivAnalymuFCHT}

To derive ${\mu}_T^{\FCHT}(t)$, we begin by plugging
$\rho_{\i\f}^{\FCHT}$ from Eq.~(\ref{eq:rhos_defns}) into
$\mu_T^{\FCHT}(t)$ in Eq.~(\ref{eq:mus_defns_FCHT}) such that
\begin{equation}
  \label{eq:muFCHTinterm-1}
  \begin{aligned}  
    \mu_T^{\FCHT}(t) &=
    2\vec{\gamma}^{\intercal}(\mathbf{R}_{0})\cdot\grad^{\intercal}\vec{\gamma}(\mathbf{R}_{0})
    \cdot \tilde{\bm{\mu}}_T^{\FCHT}(t),
  \end{aligned}
\end{equation}
where
\begin{equation}
  \label{eq:muFCHTinterm-2}
  \tilde{\bm{\mu}}_T^{\FCHT}(t) 
  =
  \sum_{\i,\f}
  \frac{1}{Z_{\I}} 
  \bra{\chi_{\i}}\ket{\chi_{\f}}
  \mel{\chi_{\f}}{\hat{\mathbf{R}}-\mathbf{R}_{0}}{\chi_{\i}}e^{-iE_{\f}t}e^{-iE_{\i}\tau}.
\end{equation}
 
We obtain a more compact expression for
$\tilde{\bm{\mu}}_T^{\FCHT}(t)$ by first considering
\begin{equation}
  \sum_{\f} \bra{\chi_{\f}}\ket{\chi_{\i}}\mel{\chi_{\i}}{\left(\hat{\mathbf{R}}-\mathbf{R}_{0}\right)}{\chi_{\f}} e^{-iE_{\f}t}
  =
  \mel{\chi_{\i}}{\left(\hat{\mathbf{R}}-\mathbf{R}_{0}\right)e^{-i\hat{H}_{\F}t}}{\chi_i},
\end{equation}
such that
\begin{equation}
  \begin{aligned}
    \tilde{\bm{\mu}}_T^{\FCHT}(t) &= \frac{1}{Z_{\I}} \sum_{\i}
    e^{-iE_{\i}\tau}
    \mel{\chi_{\i}}{\left(\hat{\mathbf{R}}-\mathbf{R}_{0}\right)e^{-i\hat{H}_{\F}t}}{\chi_i} \\
    &= \int\int \frac{1}{Z_{\I}} \sum_{\i} e^{-iE_{\i}\tau}
    \bra{\chi_{\i}}\ket{\mathbf{R}}
    \mel{\mathbf{R}}{\left(\hat{\mathbf{R}}-\mathbf{R}_{0}\right)e^{-i\hat{H}_{\F}t}}{\mathbf{R'}}
    \bra{\mathbf{R'}}\ket{\chi_{\i}}
    d\mathbf{R} d\mathbf{R'} \\
    &= \int\int \frac{1}{Z_{\I}} \left(\int
      \mel{\mathbf{R}}{\left(\hat{\mathbf{R}}-\mathbf{R}_{0}\right)}{\mathbf{R}''}\mel{\mathbf{R}''}{e^{-i\hat{H}_{\F}t}}{\mathbf{R}'}
      d\mathbf{R}'' \right)
    \mel{\mathbf{R'}}{e^{-i\hat{H}_{\I}t}}{\mathbf{R}} d\mathbf{R}
    d\mathbf{R'}.
  \end{aligned}
\end{equation}
Given the freedom to Taylor expand $\mathbf{\gamma}(\mathbf{R})$ from
arbitrary points, one can choose $\mathbf{R}_{0}=0$ and recalling
$\mel{\mathbf{R}}{\hat{\mathbf{R}}}{\mathbf{R}''}=\mathbf{R}\delta(\mathbf{R}-\mathbf{R}'')$
we have
\begin{equation}
  \label{eq:muFCHT_temp_1}
  \begin{aligned}
    \tilde{\bm{\mu}}_T^{\FCHT}(t) &= \int\int \mathbf{R}
    \mel{\mathbf{R}}{e^{-i\hat{H}_{\F}t}}{\mathbf{R}'} \frac{
      \mel{\mathbf{R'}}{e^{-i\hat{H}_{\I}t}}{\mathbf{R}} } { Z_{\I} }
    d\mathbf{R} d\mathbf{R'}. \\
  \end{aligned}
\end{equation}
Since $\mathbf{R}$ is an internal nuclear coordinate, it can be
replaced in Eq.~(\ref{eq:muFCHT_temp_1}) by the normal modes
$\mathbf{Q}$ of relevant PESs computed at their minima. Since the
product between the two propagators divided by the partition function
was determined in Eq.~(\ref{eq:muFC_final}), we can express
Eq.~(\ref{eq:muFCHT_temp_1}) as follows
\begin{equation}
  \tilde{\bm{\mu}}_T^{\FCHT}(t) = 
  \det(\mathbf{\Omega}_{\I})\sqrt{\frac{\det(\mathbf{A}_{\F})}{(2\pi i)^{N}}}
  \int\int \mathbf{Q}_{\F}
  e^{\left(
      \frac{i}{2}
      \begin{bmatrix}
	\mathbf{Q}_{\F} & \mathbf{Q}_{\F}'
      \end{bmatrix}
      \mathbf{D}
      \begin{bmatrix}
	\mathbf{Q}_{\F} \\
	\mathbf{Q}_{\F}'
      \end{bmatrix}
      +
      i
      \mathbf{E}^{\intercal}
      \begin{bmatrix}
	\mathbf{Q}_{\F} \\
	\mathbf{Q}_{\F}'
      \end{bmatrix}
      +
      i
      {F}
    \right)} 
  d\mathbf{Q}_{\F}d\mathbf{Q}_{\F}'.\\
\end{equation}
To compute the integral above, we denote
$\mathbf{E^{\intercal}}=\begin{bmatrix} \mathbf{E}_{1}^{\intercal} &
  \mathbf{E}_{2}^{\intercal} \end{bmatrix}$ where
$\mathbf{E}_{1}=\mathbf{K}^{\intercal}(\mathbf{B}_{\I}-\mathbf{A}_{\I})\mathbf{J}_{\I\F}$
and
$\mathbf{E}_2=\mathbf{K'}^{\intercal}(\mathbf{B}_{\I}-\mathbf{A}_{\I})\mathbf{J}_{\I\F}$
such that

\begin{equation}
  \begin{aligned} 
    \tilde{\bm{\mu}}_T^{\FCHT}(t) &=
    \det(\mathbf{\Omega}_{\I})\sqrt{\frac{\det(\mathbf{A}_{\F})}{(2\pi
        i)^{N}}} \left(-i\right) \frac{\partial}{\partial\mathbf{E}_1}
    \int\int e^{\left( \frac{i}{2}
	\begin{bmatrix}
          \mathbf{Q}_{\F} & \mathbf{Q}_{\F}'
	\end{bmatrix}
	\mathbf{D}
	\begin{bmatrix}
          \mathbf{Q}_{\F} \\
          \mathbf{Q}_{\F}'
	\end{bmatrix}
	+ i \mathbf{E}^{\intercal}
	\begin{bmatrix}
          \mathbf{Q}_{\F} \\
          \mathbf{Q}_{\F}'
	\end{bmatrix}
	+ i {F} \right)}
    d\mathbf{Q}_{\F}d\mathbf{Q}_{\F}'\\
    &= \det(\mathbf{\Omega}_{\I}) \sqrt{\frac{(-2\pi
        i)^{N}\det(\mathbf{A}_{\F})}{\det(\mathbf{D})}}
    \left(-i\right) \frac{\partial}{\partial\mathbf{E}_1}
    e^{\left(i\left( -\frac{\mathbf{E}^{\intercal}\mathbf{D}^{-1}\mathbf{E}}{2} + {F} \right) \right)} \\
    &= \left. \det(\mathbf{\Omega}_{\I}) \sqrt{\frac{(-2\pi
          i)^{N}\det(\mathbf{A}_{\F})}{\det(\mathbf{D})}}
      \left(-i\right) e^{\left(i\left(
            -\frac{\mathbf{E}^{\intercal}\mathbf{D}^{-1}\mathbf{E}}{2}
            + {F} \right) \right)} \left(\frac{-i}{2}\right)
      \left(\mathbf{D}^{-1}+\left(\mathbf{D}^{-1}\right)^{\intercal}\right)\mathbf{E} \right]_{[1:N]} \\
    &= -\left. \frac{\mu_T^{\FC}(t)}{2}
      \left(\mathbf{D}^{-1}+\left(\mathbf{D}^{-1}\right)^{\intercal}\right)\mathbf{E}
    \right]_{[1:N]},
  \end{aligned}
\end{equation}
where the notation $\mathbf{u}=\left.\mathbf{v}\right]_{[1:N]}$
implies that $\mathbf{u}$ and $\mathbf{v}$ are N and 2N dimensional
column arrays respectively and that $\mathbf{u}$ is the upper half of
$\mathbf{v}$ (i.e. comprised of the array elements of $\mathbf{v}$
indexed from 1 to N).

\section{Derivation of the analytical expression of
  ${\bm{\mu}}^{\HT}(t)$}
\label{subsec:derivAnalymuHT}

To derive $\mu_T^{\HT}(t)$, we begin by plugging
$\rho_{\i\f}^{\HT}$ from Eq.~(\ref{eq:rhos_defns}) into
${\mu}_T^{\HT}(t)$ from Eq.~(\ref{eq:mus_defns_HT}),
\begin{equation}
  \label{eq:muHTinterm-1}
  \mu_T^{\HT}(t) = \sum_{\i,\f}\frac{1}{Z_{\I}}
  \mel{\chi_{\i}}{\left(\hat{\mathbf{R}}-\mathbf{R}_{0}\right)^{\intercal}}{\chi_{\f}}
  \grad\vec{\gamma}^{\intercal}
  \grad^{\intercal}\vec{\gamma}
  \mel{\chi_{\i}}{\left(\hat{\mathbf{R}}-\mathbf{R}_{0}\right)}{\chi_{\f}} e^{-iE_{\i}t} e^{-iE_{\f}t}.
\end{equation}
The previous expression may be reorganized by considering that given a
vector $\mathbf{v}$ and a matrix $\mathbf{M}$,
\begin{equation}
  \mathbf{v}^{\intercal} \mathbf{M} \mathbf{v} = \sum_{\xi\varepsilon} \mathbf{v}_{\xi}\mathbf{M}_{\xi\varepsilon}\mathbf{v}_{\varepsilon} = \sum_{\xi\varepsilon} 
  \lbrack
  \mathbf{M} \circ \left(\mathbf{v}\mathbf{v}^{\intercal}\right)
  \rbrack_{\xi\varepsilon}
\end{equation}
where $\circ$ denotes the Hadamard product. Consequently, we may
reorganize Eq.~(\ref{eq:muHTinterm-1}) as
\begin{equation}
  \label{eq:muHTinterm-2}
  \begin{aligned}
    \mu_T^{\HT}(t) &= \sum_{\xi\varepsilon} \Bigg \lbrack
    \lbrack\grad\vec{\gamma}^{\intercal}
    \grad^{\intercal}\vec{\gamma}\rbrack \circ
    \tilde{\bm{\mu}}_T^{\HT}(t) \Bigg \rbrack_{\xi\varepsilon},
  \end{aligned}
\end{equation}
where
\begin{equation}
  \label{eq:muHTinterm-3}
  \tilde{\bm{\mu}}_T^{\HT}(t)
  =
  \sum_{\i,\f} \frac{e^{-iE_{\i}\tau}e^{-iE_{\f}t}}{Z_{\I}}
  \Big \lbrack 
  \mel{\chi_{\f}}{\left( \hat{\mathbf{R}} - \mathbf{R}_{0} \right)}{\chi_{\i}}  
  \mel{\chi_{\i}}{\left( \hat{\mathbf{R}} - \mathbf{R}_{0} \right)^{\intercal}}{\chi_{\f}}
  \Big \rbrack
  \Bigg \rbrack
\end{equation}
is a square matrix whose axes have dimensions equal to the number of
vibrational degrees of freedom. Eq.~(\ref{eq:muHTinterm-3}) can be
rendered more compact when using
\begin{equation}
  \sum_{\f}
  e^{-i\hat{E}_{\f}t} 
  \mel{\chi_{\i}}{ \left( \hat{\mathbf{R}} - \mathbf{R}_{0} \right) }{\chi_{\f}}
  \mel{\chi_{\f}}{ \left( \hat{\mathbf{R}} - \mathbf{R}_{0} \right)^{\intercal} }{\chi_{\i}}
  =
  \mel{\chi_{\i}}{\left( \hat{\mathbf{R}} - \mathbf{R}_{0} \right) e^{-i\hat{H}_{\F}t} \left( \hat{\mathbf{R}} - \mathbf{R}_{0} \right)^{\intercal}}{\chi_{\i}},
\end{equation}
so that
\begin{equation}
  \label{eq:muHTinterm-4}
  \begin{aligned}
    \tilde{\bm{\mu}}_T^{\HT}(t) &= \sum_{\i,\f}
    \frac{e^{-iE_{\i}\tau}}{Z_{\I}}
    \mel{\chi_{\i}}{\left( \hat{\mathbf{R}} - \mathbf{R}_{0} \right) e^{-i\hat{H}_{\F}t} \left( \hat{\mathbf{R}} - \mathbf{R}_{0} \right)^{\intercal}}{\chi_{\i}}, \\
    &= \int\int \frac{1}{Z_{\I}}
    \mel{\mathbf{R'}}{e^{-i\hat{H}_{\I}\tau}}{\mathbf{R}}
    \mel{\mathbf{R}}{\left( \hat{\mathbf{R}} - \mathbf{R}_{0} \right)
      e^{-i\hat{H}_{\F}t} \left( \hat{\mathbf{R}} - \mathbf{R}_{0}
      \right)^{\intercal}}{\mathbf{R'}} d\mathbf{R}d\mathbf{R'}.
  \end{aligned}
\end{equation}
Given the freedom to Taylor expand $\mathbf{\gamma}(\mathbf{R})$ from
any point, one can choose $\mathbf{R}_{0}=0$ thereby simplifying
Eq.~(\ref{eq:muHTinterm-4}) into
\begin{equation}
  \label{eq:muHT_intm1}
  \begin{aligned}
    \tilde{\bm{\mu}}_T^{\HT}(t) &= \int\int \frac{1}{Z_{\I}}
    \mel{\mathbf{R'}}{e^{-i\hat{H}_{\I}\tau}}{\mathbf{R}}
    \mel{\mathbf{R}}{\hat{\mathbf{R}} e^{-i\hat{H}_{\F}t}
      \hat{\mathbf{R}}^{\intercal}}{\mathbf{R'}}
    d\mathbf{R}d\mathbf{R'}.
  \end{aligned}
\end{equation}
In the coordinate representation, the term on the right-hand side in
the equation above becomes
\begin{equation}
  \begin{aligned}
    \mel{\mathbf{R}}{\hat{\mathbf{R}} e^{-i\hat{H}t}
      \hat{\mathbf{R}}^{\intercal}}{\mathbf{R'}} &= \int\int
    \mel{\mathbf{R}}{\hat{\mathbf{R}}}{\mathbf{R''}}
    \mel{\mathbf{R''}}{e^{-i\hat{H}t}}{\mathbf{R'''}}
    \mel{\mathbf{R'''}}{\hat{\mathbf{R}}^{\intercal}}{\mathbf{R'}}
    d\mathbf{R''}d\mathbf{R'''} \\
    &= \int\int \delta({\mathbf{R}-\mathbf{R''}}) \mathbf{R''}
    \mel{\mathbf{R''}}{e^{-i\hat{H}t}}{\mathbf{R'''}}
    \mathbf{R'''}^{\intercal} \delta({\mathbf{R'''}-\mathbf{R'}})
    d\mathbf{R''}d\mathbf{R'''} \\
    &=
    \mathbf{R}\mel{\mathbf{R}}{e^{-i\hat{H}t}}{\mathbf{R'}}\mathbf{R'}^{\intercal}
  \end{aligned}
\end{equation}
allowing Eq.~(\ref{eq:muHT_intm1}) to become
\begin{equation}
  \label{eq:mu_HT_temp_1}
  \tilde{\bm{\mu}}_T^{\HT}(t) 
  = 
  \int\int 
  \frac{
    \mel{\mathbf{R'}}{e^{-i\hat{H}_{\I}\tau}}{\mathbf{R}}
  }
  {
    Z_{\I}
  }
  \mathbf{R}\mel{\mathbf{R}}{e^{-i\hat{H}_{\F}t}}{\mathbf{R'}}\mathbf{R'}^{\intercal}
  d\mathbf{R}d\mathbf{R'}.
\end{equation}

Like in Appendix~\ref{subsec:derivAnalymuFCHT}, we can replace
$\mathbf{R}$ by the normal modes $\mathbf{Q}$ and use the result of
Eq.~(\ref{eq:muFC_final}) to express Eq.~(\ref{eq:mu_HT_temp_1}) as
follows
\begin{equation}
  \begin{aligned}
    \tilde{\bm{\mu}}_T^{\HT}(t) &= \int\int
    \mathbf{Q}_{F}\mathbf{Q'}_{F}^{\intercal}
    \left|\mathbf{\Omega}_{\I}\right|\sqrt{\frac{\det(\mathbf{A}_F)}{(2\pi
        i)^{N}}} e^{\left( -\frac{i}{2}
	\begin{bmatrix}
          \mathbf{Q}_{\F} & \mathbf{Q}_{\F}'
	\end{bmatrix}
	\mathbf{D}
	\begin{bmatrix}
          \mathbf{Q}_{\F} \\
          \mathbf{Q}_{\F}'
	\end{bmatrix}
	+ i \mathbf{E}^{\intercal}
	\begin{bmatrix}
          \mathbf{Q}_{\F} \\
          \mathbf{Q}_{\F}'
	\end{bmatrix}
	+ i {F} \right)} d\mathbf{Q}_{\F}d\mathbf{Q}_{\F}'
    \\
    &=
    -\frac{\partial}{\partial \mathbf{E}_{1}}\left(\frac{\partial}{\partial \mathbf{E}_{2}}\right)^{\intercal}\tilde{\mu}_T^{\FC}(t) \\
    &= -\tilde{\mu}_T^{\FC}(t) \Big \lbrack \big \lbrack
    \left(\mathbf{d}_{11}+\mathbf{d}_{11}^{\intercal}\right)\mathbf{E}_{1}
    + \left(\mathbf{d}_{12} +
      \mathbf{d}_{12}^{\intercal}\right)\mathbf{E}_{2} \big \rbrack
    \big \lbrack \mathbf{E}_{2}^{\intercal}\left(
      \mathbf{d}_{22}+\mathbf{d}_{22}^{\intercal} \right) +
    \mathbf{E}_{1}^{\intercal}\left( \mathbf{d}_{21}^{\intercal} +
      \mathbf{d}_{12} \right) \big \rbrack + i\left(
      \mathbf{d}_{12}^{\intercal} + \mathbf{d}_{12} \right) \Big
    \rbrack,
  \end{aligned}
\end{equation}
where we have defined
\begin{equation}
  \begin{bmatrix}
    \bm{d}_{11} & \bm{d}_{12} \\
    \bm{d}_{21} & \bm{d}_{22}
  \end{bmatrix}
  =
  \frac{\mathbf{D}^{-1}}{2}.
\end{equation}

\section{The time-discontinuity of propagators from the path integral
  (PI) formalism}
\label{subsec:time-discon-PI}

The origin of the discontinuity in propagators is well documented and
can be found in many textbooks~\cite{tannor2007introduction,
  kleinert2009path, schulman2012techniques}. Within the PI
formulation, the propagator of one-dimension system is given as
$\mel{q}{e^{-i\hat{H}t'}}{q'} = \int_{0}^{t'} e^{iS[\mathbf{q}]}
\mathcal{D}\mathbf{q}$ where $S\left[\mathbf{q}\right]$ denotes the
action of a trajectory $\mathbf{q}=\mathbf{q}(t)$,
$\int_{0}^{t'} \mathcal{D}\mathbf{q}$ denotes integration over all
trajectories with $\mathbf{q}(0) = q$ and $\mathbf{q}(t') = q'$.

By applying the semiclassical approximation, which consists in Taylor
expanding $S[\mathbf{q}]$ around the classical trajectory
$\mathbf{q}_{c}$, the propagator can be approximated by the following
factorized form

\begin{equation}
  \label{eq:factorKernel}
  \mel{q}{e^{-i\hat{H}t'}}{q'} \approx e^{iS[\mathbf{q}_{c}]}\int_{0}^{t'}  e^{\frac{i}{2}\delta^{2} S[\delta \mathbf{q}]} \mathcal{D}\delta\mathbf{q},
\end{equation}
where $\int \mathcal{D}\delta\mathbf{q}$ denotes integration over
variations $\delta\mathbf{q} = \delta\mathbf{q}(t)$, which are subject
to boundary conditions $\delta\mathbf{q}(0) = 0$ and
$\delta\mathbf{q}(t') = 0$. The integral term in
Eq.~(\ref{eq:factorKernel}) is often referred to as the
\textit{fluctuation factor}~\cite{kleinert2009path} and in the case of
the harmonic oscillator it coincides with the pre-exponential factor
in Eq.~(\ref{eq:Mehler_kernel_1D}), the root of the
time-discontinuity. The second functional derivative of the action
\begin{equation}
  \Lambda 
  = 
  \frac{\delta^{2}S\left[\bm{q}_{c}\right]}{\delta\bm{q} \delta\bm{q'}} 
  = -m\frac{d^2}{dt^2}-\frac{\partial^2 V\left[\bm{q}_{c}\right]}{\partial\bm{q}^{2}}, 
\end{equation}
is needed to evaluate
the second variation $\delta^2 S[\delta\bm{x}]$.

To do so, a common approach~\cite{feynman2010quantum} consists in
expanding $\delta\bm{q}$ in the basis of the eigenfunctions of
$\Lambda$, $\Lambda u_{n}(t) = \lambda_{n} u_{n}(t)$ subject to
boundary conditions $u_{n}(0) = u_{n}(t') = 0$. In the case of the
harmonic oscillator, this amounts to a Fourier expansion
\begin{equation}
  \delta\mathbf{q}(t)=  \sum_{n=1}^{\infty} a_{n} u_{n}(t)= \sum_{n=1}^{\infty} a_{n} \sqrt\frac{2}{t'}\sin(\frac{n\pi t}{t'}), 
\end{equation}
which results in the fluctuation factor being proportional to
$\prod_{n=1}^{\infty} \lambda_{n}^{{-1}/{2}} $ where
$\lambda_{n}=m\left[\left(\frac{n\pi}{t'}\right)^2-w^{2}\right]$. The
term $\lambda_{n}$ can be interpreted to be the curvature of
$S[\mathbf{q}]$ near $\mathbf{q}_c$ along the direction $u_{n}(t)$.
All $\lambda_{n}$ starts off by being positive before becoming
negative after their half-periods. The classical harmonic trajectory
minimizes the action when $t'<\frac{\pi}{w}$ and becomes a saddle
point when $t'>\frac{\pi}{w}$. Ultimately, this switch is the reason
behind the discontinuity of $\bar{\mu}_T^{\FC}(t)$.

\bibliography{diverg_free_acs}

\makeatletter
\setlength\acs@tocentry@width{1.75in}
\setlength\acs@tocentry@height{3.25in}
\makeatother

\renewcommand\tocentryname{TOC Graphic}

\begin{tocentry}
  \begin{center}
    \includegraphics{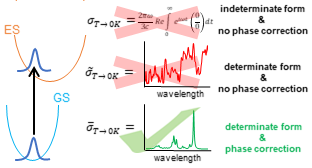}
  \end{center}
\end{tocentry}

\end{document}